\documentclass[%
 reprint,
 amsmath,amssymb,
 aps,
float,
]{revtex4-2}

\usepackage{graphicx}
\usepackage{dcolumn}
\usepackage{bm}

\usepackage{epstopdf}
\usepackage{xcolor}
\usepackage{soul}
\AppendGraphicsExtensions{.tif}

\begin{document}

\title{Physically Intuitive Anisotropic Model of Hardness}

\author{Faridun N. Jalolov}
\affiliation{Skolkovo Institute of Science and Technology, Bolshoy Boulevard 30, bld. 1, Moscow 121205, Russia}

\author{Alexander G. Kvashnin*}
\affiliation{Skolkovo Institute of Science and Technology, Bolshoy Boulevard 30, bld. 1, Moscow 121205, Russia}

\email{A.Kvashnin@skoltech.ru}

\date{\today}

\begin{abstract}
The hardness of materials plays an important role in material design.
There are numerous experimental methods to measure the hardness of materials, but theoretical prediction of hardness is challenging.
By investigating the correlation between hardness and the elastic properties of materials, namely shear and bulk moduli, the pressure derivative of bulk modulus, we have constructed a simple and physically intuitive hardness model.
By introducing the spatial variation of the shear modulus, it is possible to predict the hardness anisotropy of materials to define the minimum and maximum values of hardness possessed by a particular material.
Furthermore, by using the equation of states to define the pressure derivative of the bulk modulus, it is possible to determine the temperature dependencies of hardness for given materials.
All quantities in the model can be obtained directly from accurate first-principles calculations or from experiments, making it suitable for practical applications.

\end{abstract}

\maketitle

\section{Introduction}
\label{sec:intro}
Hardness is an important property of materials that defines their ability to resist indentation or scratching by another, harder material.
The determination of hardness is traditionally based on the experimental method of indenting the test sample with a harder material,
usually diamond.
In this case, the hardness is determined from the relationship between the maximum indentation force and the impression left on the test specimen.

From another perspective, modern industries require new hard and superhard materials with improved mechanical properties compared to traditional materials.
One solution to this problem lies in the application of modern computational techniques to high-throughput screening for materials with improved properties. 
Computational techniques are now sufficiently developed to accurately predict the structure and properties of various compounds and materials \cite{kvashnin_cmd, oganov2, oganov3}.
However, it is not only important to predict the structure of a material, but also to accurately calculate its mechanical properties, such as hardness, which are required for targeted experimental synthesis of a material with predefined properties.

Currently, a number of theoretical empirical models are used to calculate the hardness of materials using atomistic modelling, which allows the prediction of hardness based on the crystal structure and some physical properties.
Thus, the empirical models proposed in Refs. \cite{Gaomodel, Simunek, LyakhovOganov} are used to predict the hardness of covalent and ionic compounds based on the strength of chemical bonds, the degree of ionicity \cite{Gaomodel, Simunek} and the electronegativity of crystals \cite{LyakhovOganov}.
A number of other models allow the prediction of hardness from the elastic moduli of materials.
This is an accurate and efficient method as all the properties for the models can be easily obtained from DFT or from experiments.  
Each of the existing model was built on the basis of approximation of available experimental measurements of hardness of different materials. Each of the hardness models has both advantages and disadvantages, and it describes well the average hardness of a large number of known compounds\cite{Kvashnin2019}.
It should be noted that almost none of the existing hardness models take into account the anisotropy of the crystal.
For example, the hardness of diamond measured in different crystallographic directions can differ by up to 50\% \cite{kraus1939variation, blank1999mechanical}. 
Recently, Podryabinkin \textit{et al.} \cite{podryabinkin2022nanohardness} proposed an alternative approach to calculating hardness that differs from the empirical models.
This method uses first-principles calculations and machine learning capabilities to explicitly simulate the nanoindentation of a material with different surfaces.
Although this method is highly accurate, it is too computationally expensive for high-throughput screening.

Here we propose a simple and accurate hardness model based on material properties such as shear modulus and pressure derivative of bulk modulus.
Both properties can be derived from experiments or from atomistic modelling.
Experimentally measured hardness values were used to fit the model, which was found to be quite simple but physically interpretable.
A data set of 103 compounds is generated to check the correlations of the proposed model with known hardness models such as Chen's model \cite{chen_modeling_2011} and Teter's model \cite{teter1998computational}. 
The use of the shear modulus as a characteristic for the model allowed us to simulate the spatial distribution of hardness, while the presence of the equation of states, i.e. the pressure derivative of the bulk modulus, gives the possibility to simulate the temperature dependence of hardness, which is compared with available experimental data for ReB$_2$ and B$_4$C. 


\section{Methods}

\subsection{Computational Details}
As mentioned above for isotropic materials, the shear modulus is scalar and does not depend on the orientation of the material.
However, all crystals are anisotropic materials and the shear modulus can vary with direction and the rotation tensor, as described by Euler angles, plays a crucial role in its determination.
To understand how the shear modulus depends on the rotation tensor and Euler angles, it is essential to consider the transformation of the elastic tensor with respect to rotation.
When analysing elastic properties such as shear modulus in different orientations, it is useful to describe the direction of elastic strain using spherical coordinates. 
An elastic strain direction can be represented by a point on the unit sphere (unit vector) as well as by two angles. 
This direction in Cartesian coordinates corresponds to uniaxial stress or response to isotropic pressure.
We have chosen it for the calculations as the first unit vector directed along the first vector of the unit cell - $a$.
It is completely characterised by the angles $\Theta$(0,$\pi$) and $\phi$(0,2$\pi$) \cite{MARMIER20102102}.
The perpendicular direction also requires the determination of some elastic properties (e.g. shear modulus, Poisson's ratio).
This direction is defined by another vector $b$. $b$ is perpendicular to vector $a$ and can be characterised by $\chi$(0, 2$\pi$).
By definition, the vectors $a$ and $b$ are the components of the first two columns of the rotation matrix.
This is sufficient to obtain, for example, all the fourth-order components in the subvector space defined by directions 1 and 2:

\begin{equation}\label{eq:S12}
S'_{12} = S'_{1122} = r_{1j} r_{1j} r_{2k} r_{2l} S_{ijkl} = a_i a_j b_k b_l S_{ijkl}
\end{equation}
\begin{equation}\label{eq:S12}
S'_{66} = S'_{1212} = r_{1j} r_{2j} r_{1k} r_{2l} S_{ijkl} = a_i b_j a_k b_l S_{ijkl}
\end{equation}

But by scanning $\Theta$, $\phi$ and $\chi$ over the unit sphere, we can access all the components without having to consider the third unit
vector. 

The shear modulus depends on two directions (if perpendicular, this corresponds to three angles), which makes it difficult to represent graphically.
A convenient way is to consider three plots: minimum, average and maximum.
For each $\Theta$ and $\phi$, the angle $\chi$ is scanned and the minimum, average and maximum values for that direction are recorded.

The shear ratio is obtained by applying a pure shear stress in vector form

\begin{equation}\label{eq:eps}
\epsilon_{ij} = S_{ijkl} \sigma_{kl}
\end{equation}

and results in

\begin{equation}\label{eq:G}
G(\Theta, \phi, \chi) = \frac{1}{4S'_{66}(\Theta, \phi, \chi)},
\end{equation}
Thus, the presence of the shear modulus in the new hardness model makes it possible to calculate the hardness anisotropy as follows


\begin{equation}\label{eq:Hv0}
H_{V}(\Theta, \phi, \chi) \propto {G(\Theta, \phi, \chi)},
\end{equation}

The Elate module \cite{Elate} was used to calculate the shear modulus anisotropy and hardness. This module is used to calculate the shear modulus in different directions of the crystal using the loaded elasticity tensor.

We would like to include the pressure derivative of the bulk modulus ($B^{'}$) in the new hardness model.
The reason for including this property in the new model is that its values can be obtained both experimentally and through calculations based on the equation of state.
In addition, the equation of state (EOS) and $B^{'}$ are both temperature dependent and it is possible to calculate the hardness dependence at finite temperature.
The temperature variation of the pressure derivative of the bulk modulus can be calculated directly. At zero temperature we define the bulk modulus as follows

\begin{equation}\label{eq:B}
B = V_0 \cdot \frac{d^2 E_0(V)}{d V^2},
\end{equation}
then the pressure derivative defines as follow

\begin{equation}\label{eq:B'}
B' = V_0 \cdot \frac{d^3 E_0(V)}{dp dV^2},
\end{equation}
where $E_0(V)$ is the total energy of structure at the volume $V$, $V_0$ is the equilibrium volume, $p$ is the pressure at 0 K.
To take the temperature contribution into account it is necessary to have the Gibbs free energy instead of the total energy:

\begin{equation}\label{eq:Gibbs1}
G(V, T) = E_{0}(T) + F_{vib}(V,T) + pV,
\end{equation}
where $F_{vib}$ is the Helmholtz vibrational  free energy, calculated in the harmonic approximation from the following relation \cite{Kern} 

\begin{equation}\label{eq:Gibbs}
\begin{split}
F_{\text{vib}}(V, T) &= k_{B}T \int g(\omega, V) \ln\left[1 - \exp\left(\frac{\hbar\omega}{k_{B}T}\right)\right] \,d\omega \\
&\quad+ \frac{1}{2} \int g(\omega, V) \hbar\omega \,d\omega.
\end{split}
\end{equation}

where $g(\omega, V)$ is the phonon density of states at the given volume calculated by the finite displacement method as implemented in PHONOPY \cite{phonopy1, phonopy2} with forces calculated by VASP \cite{vasp1, vasp2, vasp3}; $k_{B}$ is the Boltzmann constant; and $\omega$ is the phonon frequency.
And the pressure derivative of the bulk modulus as a function of temperature is defined as follows

\begin{equation}\label{eq:B'}
B'(T) = V \cdot \frac{d^3 F(V, T)}{dp dV^2},
\end{equation}
which allows us to take into account the temperature contribution to the hardness.

\section{Results and Discussion}
\label{sec:result}

\subsection{Model and Data Set}

\label{sec:comp}
Among the various number scales of hardness, we have chosen Vickers hardness.
Vickers hardness is one of the most convenient, widely used in testing, and many experimental data are available.
It can be used for most solids, has a well-defined number scale, and the variance of the values in different experiments is often small compared to other methods. 
We would like to find a physically intuitive model of hardness based on physical quantities that are directly or indirectly related to the hardness of a material.
The experimental data for a wide variety of materials show that the shear modulus $G$, which measures the material's resistance to shape change, is correlated with hardness.
The Teter model \cite{teter1998computational} identifies the good correspondence between hardness and shear modulus by a linear dependence $H_V = 0.151 \cdot G$.
From another point of view, the pressure derivative of the bulk modulus at zero pressure ($B^{'}$) is a parameter of great physical importance in physics and materials science.

\begin{figure}[ht]
  \begin{center}
      \center{\includegraphics[width=1\linewidth]{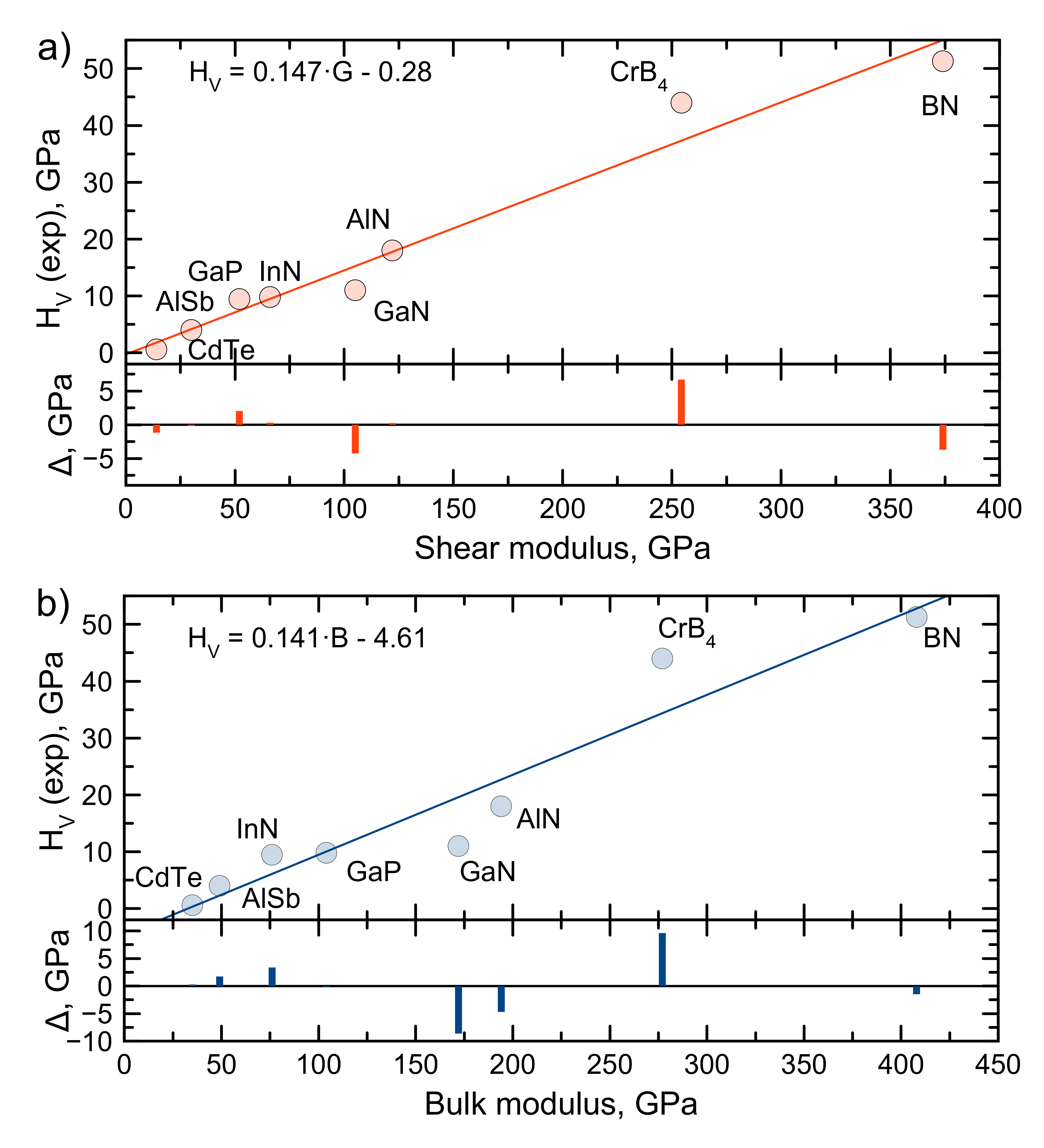}}  \\
  \caption {The correlations between experimentally measured hardness of selected compounds and a) shear, and b) bulk moduli. Bottom panels show the difference between the actual data and the linear fit.}
  \label{fig:1}
  \end{center}
\end{figure}

It is being related to a few other important thermophysical properties and besides being intimately linked to the equation of state (EOS) and it is an experimentally measurable quantity.
Furthermore, from the point of view of mechanical properties and experimental observations, the pressure derivative of the bulk modulus should be inversely proportional to the hardness. 
For example, the $B^{'}$ for cubic BN is about 3.6 \cite{knittle1989experimental} and the hardness is about 60 GPa, while for diamond it is 3.01 \cite{occelli2003properties} with the hardness of about 98 GPa.  
According to this, we suggest to propose the model that will be proportional to the shear modulus and inversely proportional to the pressure derivative of the bulk modulus. 
Proportionality to the shear modulus is also convenient to use because it is directional, so we can use this property of the shear modulus to construct the model that will depend on the spatial orientation.
As a result, we have proposed the following relationship

\begin{equation}\label{eq:Hv}
H_{V} = a \cdot \frac{G}{(B^{'})^{c}},
\end{equation}
where $B^{'}$ is a pressure derivative of the bulk modulus, $G$ is the shear modulus, $a$ and $c$ are the fitting parameters.
To fit the parameters $a$ and $c$ we have collected the database of experimentally measured hardness values, shear modulus and $B^{'}$.
Our data include a number of high symmetry ionic and covalent materials, namely AlN, CdTe, GaN, GaP, InN, Si, ZnSe, ZnTe, BN.
The numerical values of the collected experimental data are presented in Supporting Information Table S1. 
We have plotted the correlations between experimentally measured hardness with shear and bulk moduli, as shown in Figure \ref{fig:1}.
We have excluded data for diamond because its hardness is difficult to measure experimentally.
During the fitting, we determined the final formula (the parameter in the degree of $B^{'}$ was rounded to the nearest integer), which allows us to compute the hardness as follows: 

\begin{equation}\label{eq:Hv}
H_{V} = 2.5 \cdot \frac{G}{(B^{'})^{2}},
\end{equation}

As a result, we have obtained a simple and physically interpretable model with minimal fitting parameters. 
Moreover, as mentioned above, this model allows the description of the anisotropy of hardness, since the shear modulus is a spatially dependent property of a material.

\subsection{Testing the model}

To test our model and show its accuracy, we collected data from a specific dataset.
The data set was collected from the Materials Project \cite{jain2013commentary} database.
We collect 103 compounds and calculate their hardness values using Chen's \cite{chen_modeling_2011} and Teter's \cite{teter1998computational} models, spanning a range from 1 to 50 GPa.
Data for selected compounds are presented in the Supporting Information \cite{SI2024}, see Table S1.
The hardness values obtained by the Chen and Teter models are provided only as a benchmark to evaluate the accuracy and reliability of our prediction approach over a wide range of materials with different hardness. 
According to the Teter model, hardness is directly proportional to the shear modulus of a compound. 
Chen's model is more complicated compared to the Teter model \cite{teter1998computational}, but the main contribution still comes from the shear modulus \cite{chen_modeling_2011}. 
Correlations between shear and bulk moduli with Chen's and Teter's hardness values for selected compounds, as well as the correlation between Chen's and Teter's hardness, are presented in the Supporting Information \cite{SI2024}, see Figure S1.

\begin{figure}[ht]
  \begin{center}
      \center{\includegraphics[width=1\linewidth]{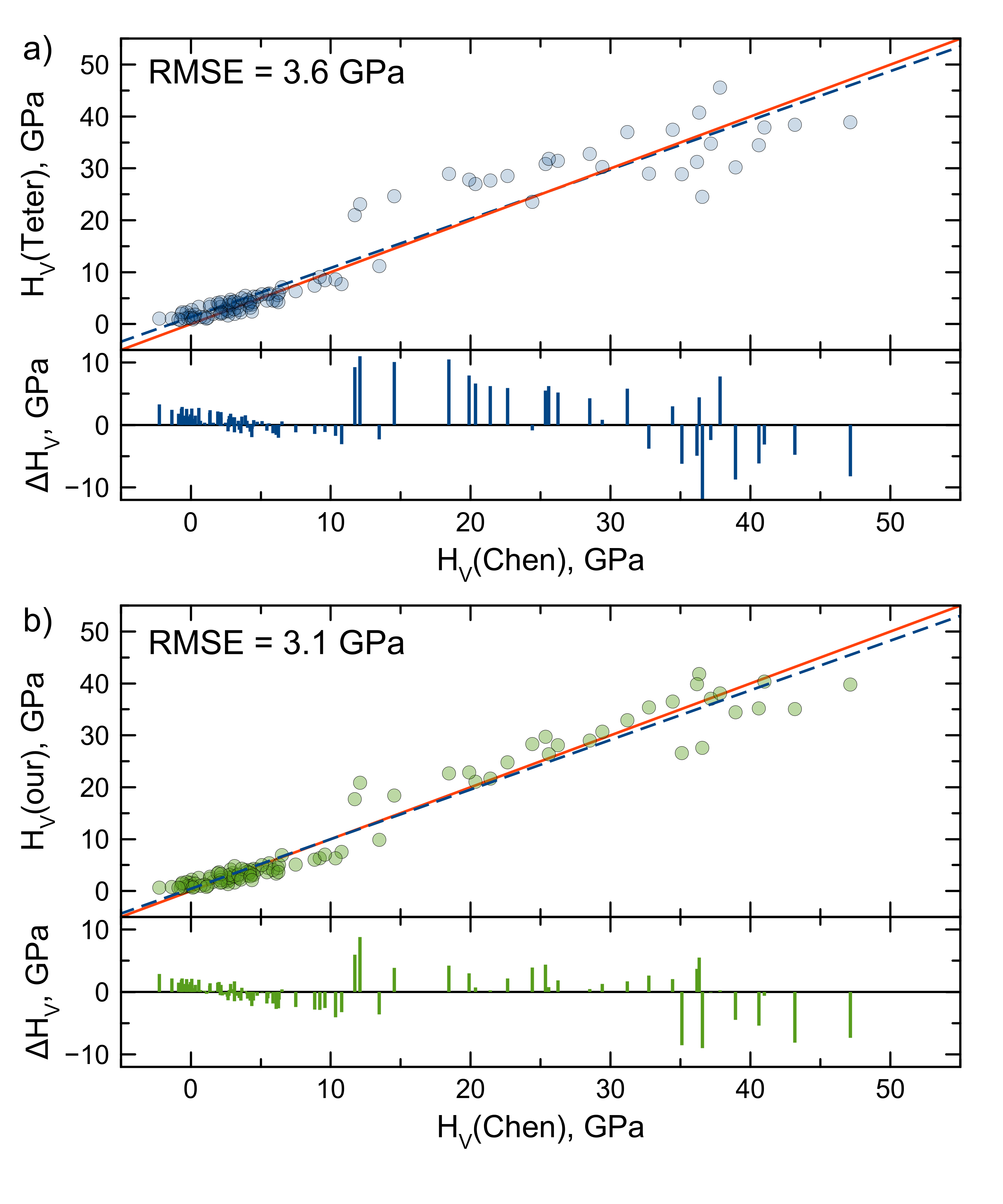}}  \\
  \caption {The correlations between Chen's model of hardness and a) Teter, b) our model of hardness calculated for a data set of 103 compounds. Red solid line corresponds to $R^2=1$, blue dashed line is a linear fit for a particular data. Bottom panels show the difference between the actual data and the linear fit.} 
  \label{fig:model_chen}
  \end{center}
\end{figure}

We perform the calculations of hardness using our model (see eq. \ref{eq:Hv}) for selected materials.
Using our data set of 103 compounds, we calculated a correlation between our model and other considered empirical models, i.e. Chen's and Teter's models.
The obtained correlations are shown in Figure \ref{fig:model_chen}.
The comparison of the predictions obtained with Teter's, Chen's and our model with experimental data is presented in the Supporting Information \cite{SI2024}, see Figure S2.
It can be seen that the root mean square error (RMSE) for our model is 3.1 GPa, while Teter's model gives 3.6 GPa with respect to Chen's hardness values. 
As can be seen, the maximum error is observed for compounds with high hardness.
The average error is not more than 5\% for all considered structures.

\subsection{Spatial distribution of hardness}

Since we have used the shear modulus as the main parameter in our model, we can predict the changes in hardness with respect to the orientation of the crystal using eq. \ref{eq:G}. 
Thus, we can obtain the following formula:

\begin{equation}\label{eq:Hv_sp}
H_{V} = 2.5 \cdot \frac{G(\Theta, \phi, \chi)}{(B^{'})^{2}}
\end{equation}

\begin{figure*}[ht]
  \begin{center}
      \center{\includegraphics[width=1\linewidth]{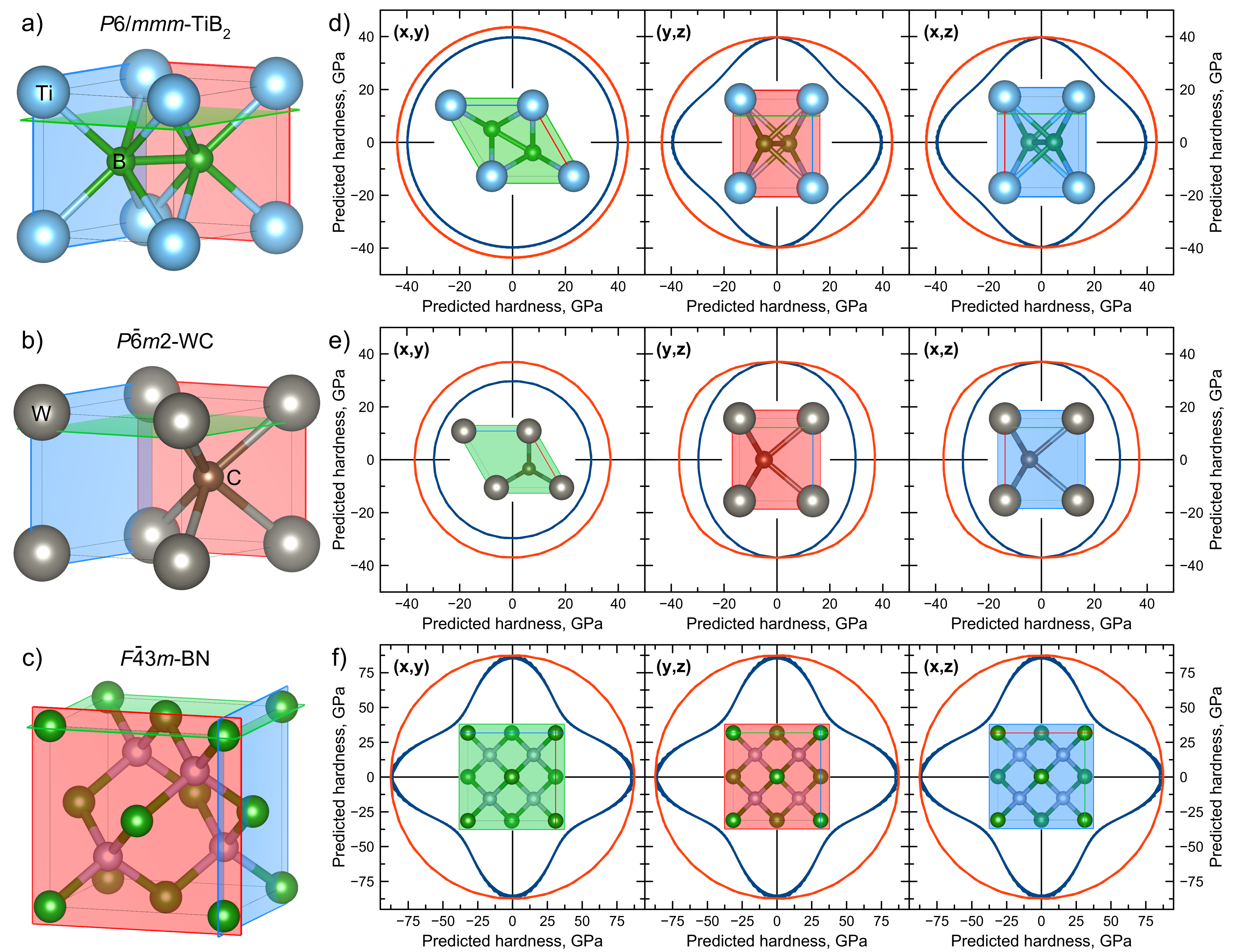}}  \\
  \caption {Crystal structures of a) TiB$_2$, b) WC, c) BN and corresponding crystal planes along which the distribution of hardness is calculated. Distribution of the calculated hardness as a function of the direction in the crystal calculated for d) TiB$_2$, e) WC and f) cubic BN. The values calculated for the two shear directions are shown by orange and blue colors.} 
  \label{fig:po}
  \end{center}
\end{figure*}

To calculate the spatial dependence of the hardness for each material, we only need to perform the calculations of $G(\Theta, \phi, \chi)$ and $B'$. 
The figure \ref{fig:po} shows the typical distribution diagrams of the calculated spatial dependence of hardness for cubic $TiB_2$, $WC$ and BN.
The dependence of the hardness on the crystal orientation is clearly visible.
For $TiB_2$, Figure \ref{fig:po}a, the hexagonal symmetry of the structure reflects the hardness distribution. 
The maximum hardness of 43.7 GPa is predicted for the $<001>$ direction.

In the case of tungsten carbide, Figure \ref{fig:po}b, its crystal structure belongs to the hexagonal $P\bar{6}\textit{m}2$ space group, which reflects the symmetry of the elastic tensor as well as all the mechanical properties. 
Here one can see the difference between the hardness distributions obtained for the (x,y) plane of WC and those including the z-direction, see figure \ref{fig:po}b.
The hardness along z-direction is 37 GPa, which is the maximum value for WC. 
The obtained values are comparable with those measured experimentally (27.4 GPa) \cite{HUANG200841} and those predicted by other models (35 GPa Mazhnik-Oganov model, 33.5 Chen's model) \cite{mazhnik, chen_modeling_2011}. 
Lee \cite{leewc} has shown the Knoop hardness values for the WC single crystal measured in different orientations with a load of 9.81 N. It has been shown that the hardness of WC is extremely anisotropic, with the hardness varying greatly depending on the crystallographic orientation of each surface as well as the orientation of the indenter itself.The Knoop hardness value for WC crystal in the (0001) direction at room temperature is about 20 GPa, while for the (1010) direction it is about 17 GPa. The values calculated by our model are 36 GPa and 27 GPa, respectively. The difference between the experimentally measured hardness and the predicted hardness is due to the relatively high load. Lower loads result in higher hardness values\cite{broitman}.

Due to the high symmetry of cubic BN ($\textit{F}$$\bar{4}$3$\textit{m}$ space group) the obtained hardness distribution is also highly symmetric with respect to crystal directions, see figure \ref{fig:po}c.
The highest value of hardness is 87.6 GPa corresponding to $<100>$ directions, see figure \ref{fig:po}a. 

All information about spatial distribution of hardness for considered compounds is collected in the Supporting Information \cite{SI2024}.

For experimentally known materials, which were used to fit the model, we have also performed the calculations of spatially dependent hardness according to the eq. \ref{eq:Hv_sp}.
This information is used to construct the correlation of the predicted minimum, maximum and average values of hardness with the experimentally measured hardness, as shown in Figure S4 in the Supporting Information \cite{SI2024}, see also references \cite{kohn1965self, hafner2008ab, YOURTCU198149, methfessel1989high,  BNBprime, perdew1996generalized, hohenberg1964inhomogeneous, TSoma, KANOUN20041601, 10.1063/1.1659349, solozhenko1994properties, dandekar1994equation, ueno1994stability} therein.

\begin{figure}[ht]
  \begin{center}
      \center{\includegraphics[width=0.95\linewidth]{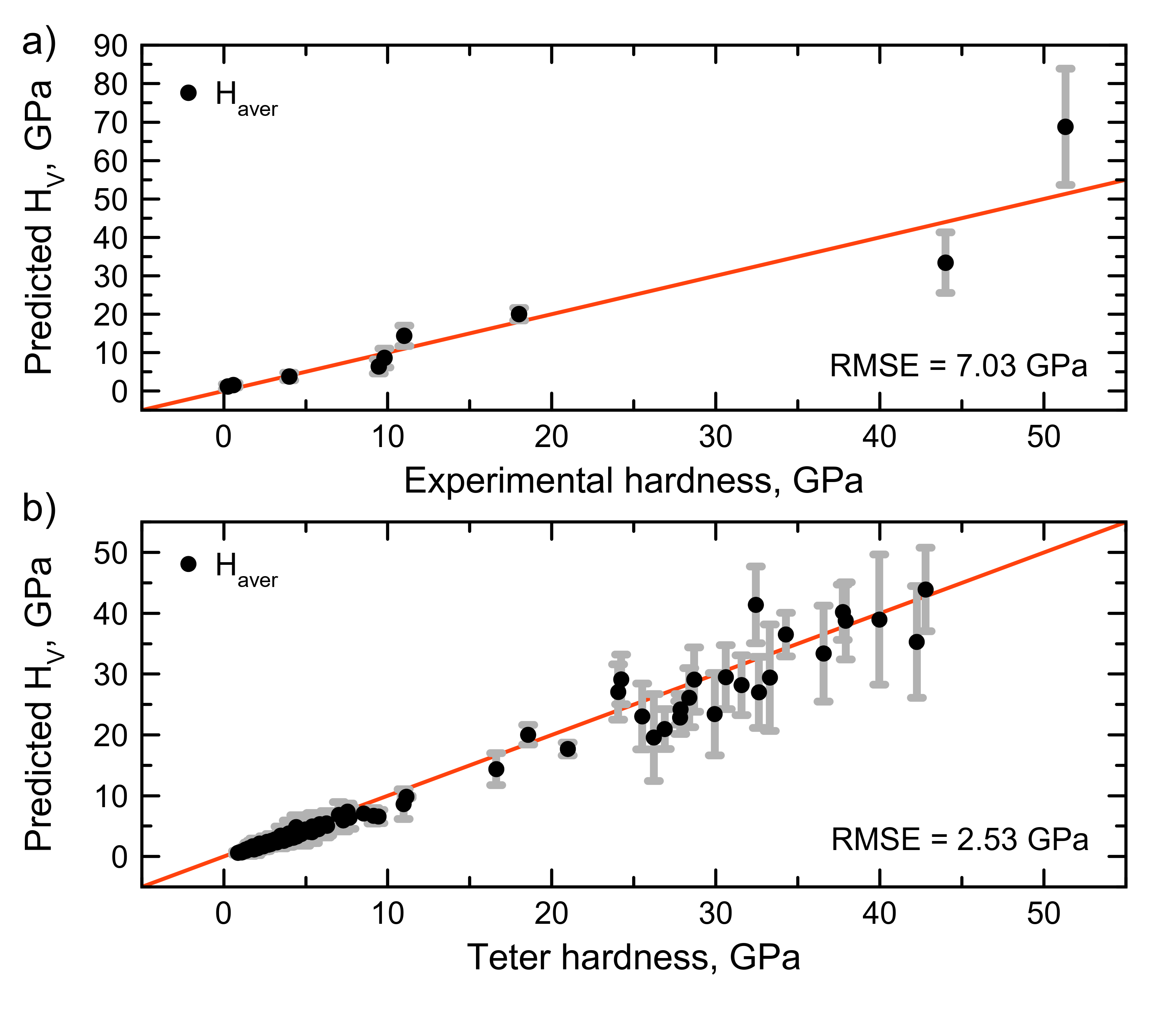}}  \\
  \caption {The correlations of the calculated average value of hardness with a) experimental data and b) the Teter model. The error bars obtained from the maximum and minimum values of hardness within our model. The red solid line shows the line $H_{exp} = H_{calc}$.}
  \label{fig:po_max_min}
  \end{center}
\end{figure}

The boundary values obtained by our model allow us to calculate the error in determining the hardness of a compound associated with crystal anisotropy compared to the average predicted value.
The comparison of the predicted hardness with the experimentally measured one is shown in figure \ref{fig:po_max_min}a.
The red solid line shows the line $H_{exp} = H_{calc}$. 
Error bars, shown in gray color, are calculated using the minimum and maximum hardness values determined by our model.  
The RMSE obtained for these data is 7.03 GPa.

Next, we calculated the hardness for 103 considered compounds.
Since there are no models available that take into account the spatial distribution of hardness, we have used the Teter model as a reference, since it does not contain any variables other than shear modulus that can influence the spatial behavior. 
The comparison of the Teter model with the average hardness values obtained from the minimum and maximum values of our model is shown in Figure \ref{fig:po_max_min}b.  
It can be seen that our hardness model is in good agreement with the experimental data and the Teter model.
Obviously, the good agreement with the Teter model comes from the linear dependence of hardness on shear modulus in both models.
However, the next important possibility of our model is to consider the temperature contribution.

\subsection{Temperature dependence of hardness}

\begin{figure*}
  \begin{center}
      \center{\includegraphics[width=0.95\linewidth]{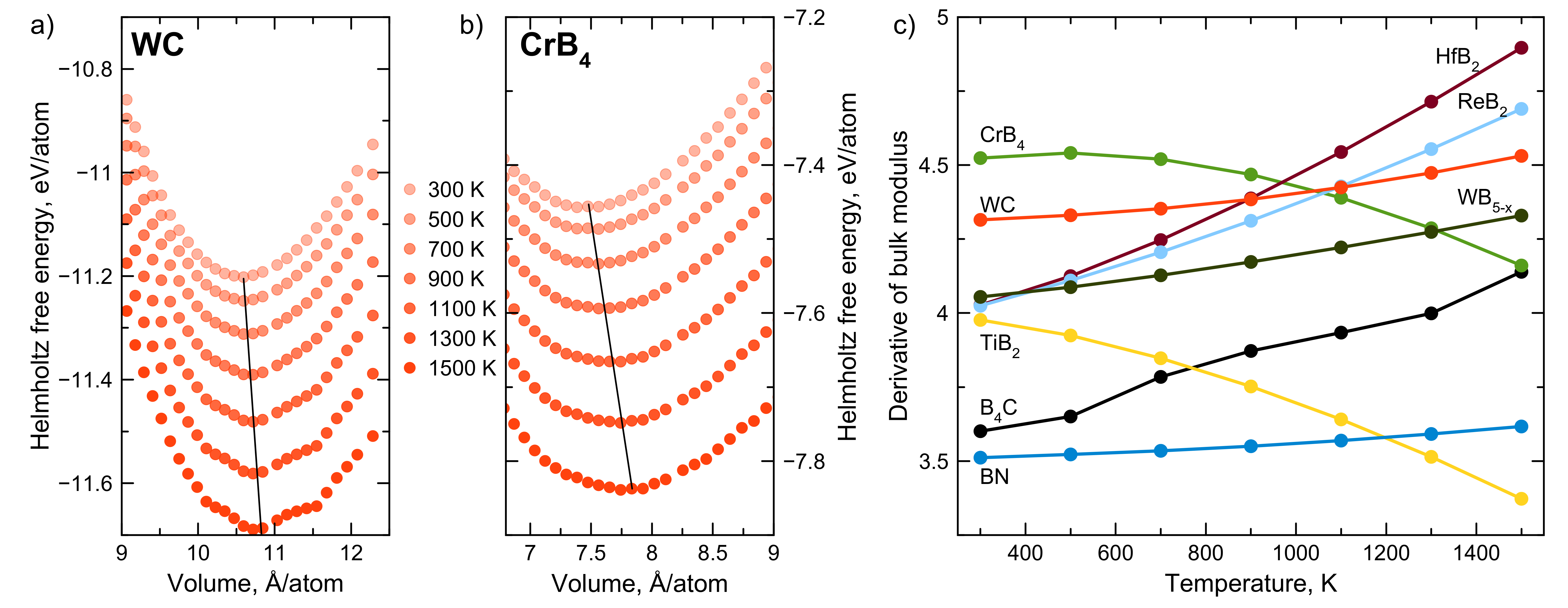}}  \\
  \caption {Calculated equations of states at different temperatures from 300 to 1500 K for a) WC and b) CrB$_4$. Black lines show the shift of the equilibrium volume with respect to temperature. c) Dependence of the pressure derivative of the bulk modulus on temperature for selected hard and superhard materials.}
  \label{fig:eos}
  \end{center}
\end{figure*}

Our model includes the pressure derivative of the bulk modulus, which is determined by calculating the equation of states (EOS). 
The equation of states itself is a temperature dependent property and can be easily calculated for any structure.
For each of the 103 structures, we perform the equation of states calculations. 
In figure \ref{fig:eos}a,b you can see the calculated equations of states for WC and CrB$_4$ as examples.
Data obtained for different temperatures from 300 to 1500 K are shown by the red gradient.
The temperature is taken into account by the Helmholtz free energy calculations as explained in the Methods section, see eq. \ref{eq:Gibbs}.
The black lines show the change of the equilibrium volume with temperature. 
For each volume point the phonon density of states is calculated and the Helmholtz free energy is evaluated. 
Thus the equation of states at different temperatures was obtained for each structure. 

\begin{figure*}
  \begin{center}
      \center{\includegraphics[width=0.95\linewidth]{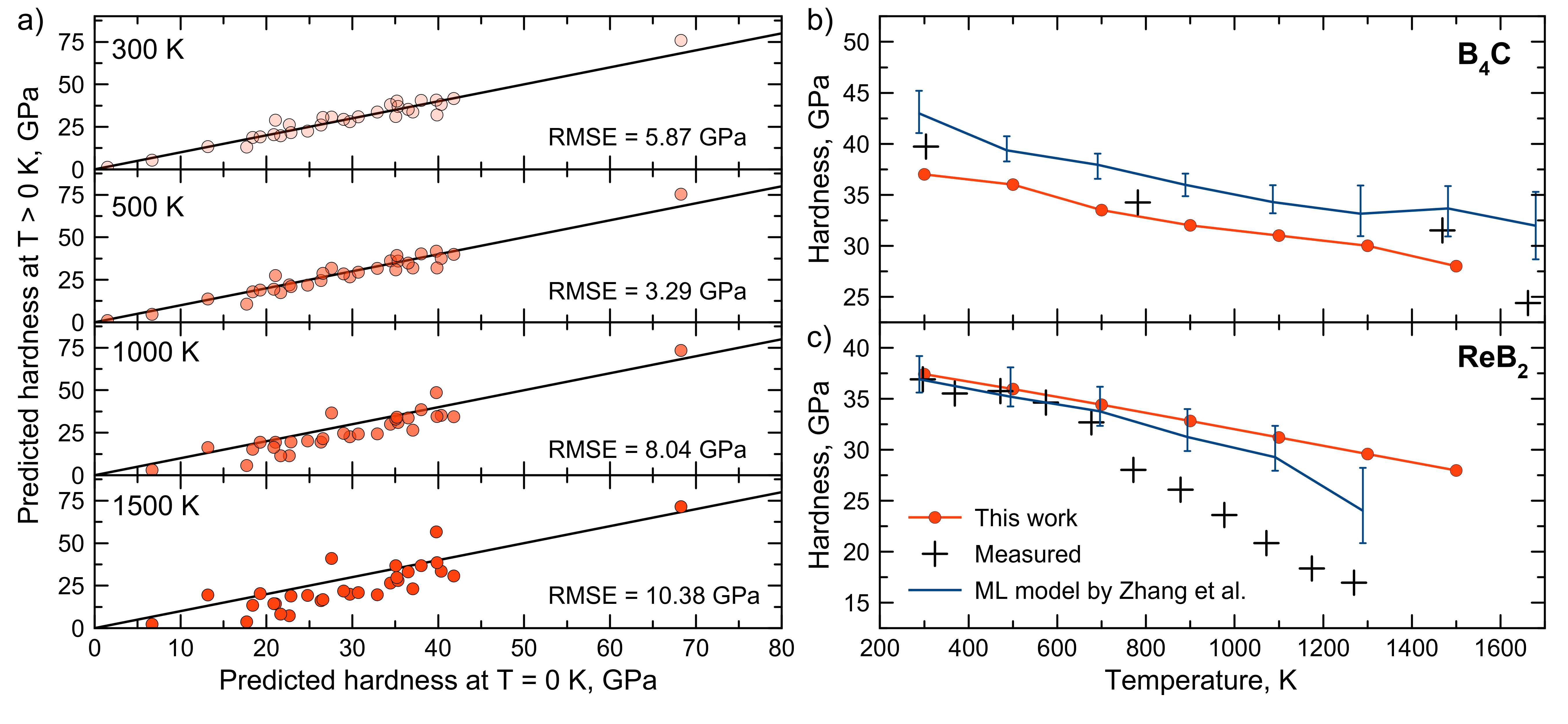}}  \\
  \caption {a) Correlations between calculated hardness at T = 0 K and finite temperatures, namely 300, 500, 1000, and 1500 K. b) Predicted hardness by our model against the experimentally measured hardness and the machine learning model from of Ref. \cite{zhang2021determining}}
  \label{fig:H_T}
  \end{center}
\end{figure*}

One can see the difference between the EOS of WC and CrB$_4$, shown in Figure \ref{fig:eos}a,b.
Increasing the temperature causes the EOS of WC to become steeper compared to CrB$_4$, where the whole dependence shifts downward in energy and the slope becomes less steep. 
This difference plays an important role in the pressure derivative of the bulk modulus calculated for selected materials and shown in Figure \ref{fig:eos}c.
It can be seen that the derivative of the bulk modulus can increase or decrease with temperature for different materials.
Such a difference in behavior is mainly related to the composition, structure and chemical bonding of the studied structures reflecting the equation of states.
However, this issue should be investigated in more detail, which was beyond the scope of this study.
For the presented hard and superhard materials, the $B^{'}$ value is in the range of 3.5 to 5 (Figure \ref{fig:eos}c).

Using these data, we have calculated the temperature dependence of the hardness $H_V(T)$ for the structures in the dataset.
The dependencies are calculated for temperatures of 300, 500, 1000, and 1500 K.
Figure \ref{fig:H_T} shows the correlation between the predicted hardness at 0 K and the hardness at the finite temperatures.
The linear trend line shows that the hardness decreases with increasing temperature.
Interestingly, the RMSE calculated for 500 K is smaller than that for 300 K, which is mainly due to the fact that cubic BN has the highest hardness and thus the highest error.
It can be seen that the increase in temperature leads to a general decrease in hardness for all considered structures compared to 0 K.

For a more detailed comparison, we selected two structures of B$_4$C and ReB$_2$, for which the experimentally measured high temperature hardness was determined in Refs. \cite{marinescu2015handbook, otani2009high}.
In addition, Zhang et al. \cite{zhang2021determining} performed predictions for these compounds using the XGBoost machine learning model.

The comparison of the experimental data and the ML predictions with our model is shown in Figure \ref{fig:H_T}b,c. 
In the case of B$_4$C we see an underestimation of the hardness with respect to the experimental values and the ML predictions. 
The underestimation is not very large, about 3 GPa, while the general slope of the dependence on temperature is reproduced very well compared to experiment. 
The difference in the predicted hardness for ReB$_2$ with the experiment is more significant. 
At low temperatures the difference is less than 1 GPa, while after 700 K the difference increases due to the change in the slope of the experimental data.
In our case, we have a linear decrease in hardness, while in the experiment the kink was observed, which cannot be reproduced by our model.
This can be related to more accurate calculations of $B^{'}$ with respect to temperature, which will influence the slope of the $B^{'}(T)$ dependence, shown in figure \ref{fig:eos}c by the blue color.

\section{Conclusion}
\label{sec:conclusion}
In this work, a new physically-inspired hardness model has been proposed and applied to a number of hard materials using information about the shear modulus and the equation of states of crystal structures.
The importance of using the shear modulus comes from its property to be calculated for different directions of the crystal structure, which allowed the calculation of the spatial dependence of hardness for a number of materials taking into account the crystal anisotropy.
The pressure derivative of the bulk modulus derived from the equation of states allowed us to consider the temperature contribution to hardness caused by the changes in the equation of states with temperature.
We have demonstrated that the model of hardness works for hard and superhard materials on the examples of ReB$_2$ and B$_4$C.
The obtained temperature dependence of hardness is in good agreement with available experimental measurements and ML predictions.
It is worth noting that all parameters in the proposed hardness model can either be calculated directly and accurately by first-principles calculations or measured experimentally, which makes the model applicable to a wide range of practical applications.

\section{Competing Interests}
The authors declare no competing financial or non-financial interests.

\begin{acknowledgements}
This work was carried out using the ElGatito and LaGatita supercomputers of the Industry-Oriented Computational Discovery lab at the Skoltech Project Center for Energy Transition and ESG.

\end{acknowledgements}

\bibliography{main.bib}

\end{document}